# Overcoming data scarcity through multi-center federated learning for organs-at-risk segmentation in pediatric upper abdominal radiotherapy

Mianyong Ding, Msc[a,b], Maximilian Knoll, MD[d], Semi Harrabi, MD[d], Martine van Grotel, MD [a], Annemieke S. Littooij, MD [a,b], Prof. Max van Noesel, MD[a,e], Prof Jens-Peter Schenk, MD[f], Prof Marry M. van den Heuvel-Eibrink, MD[a,g], Geert O. Janssens, MD[a,b], Matteo Maspero, PhD[b,c]

a Princess Máxima Centre for Pediatric Oncology, Utrecht, The Netherlands
b Department of Radiation Oncology, Imaging and Cancer Division, University Medical Centre Utrecht, Utrecht, The Netherlands
c Computational Imaging Group for MR Diagnostics & Therapy, Centre for Image Sciences, University Medical Centre Utrecht, Utrecht, The Netherlands
d University Hospital Heidelberg, Heidelberg Ion Beam Therapy Centre (HIT), Department of Radiation Oncology, Heidelberg, Germany
e Division Imaging & Cancer, University Medical Center Utrecht, Utrecht, The Netherlands
f Division of Pediatric Radiology, Department of Diagnostic and Interventional Radiology, University Hospital Heidelberg, Heidelberg, Germany; Radiology Panel, SIOP/RTSG Association
g Wilhelmina Children's Hospital-Division of CHILD HEALTH, University Medical Centre Utrecht, University of Utrecht, Utrecht, the Netherlands

## Abstract

### Background

Deep learning–based organs/structures-at-risk (OARs) auto-contouring models can improve radiotherapy workflows, but models trained on adult data often underperform in pediatric patients. Developing robust pediatric-specific models is hindered by data scarcity and fragmentation across centers. Federated learning (FL) enables privacy-preserving collaborative training without the need for data sharing. We evaluated the feasibility and performance of FL for developing pediatric-specific OAR segmentation models across two European medical centers.

### Methods

Computed tomography (CT) images from pediatric patients from Utrecht and Heidelberg with a renal tumor or abdominal neuroblastoma were retrospectively collected and locally processed. An nnU-Net–based framework segmented 19 OARs using local and FL schemes. FL was implemented with secure weight exchange on a cloud storage across institutional firewalls. Performance was assessed using the Dice similarity coefficient (DSC), 95th percentile Hausdorff distance, and mean surface distance. Robustness to patient orientation, false-positive segmentation of surgically removed kidneys, and failure cases were identified.

### Findings

A total of 310 postoperative CTs from 272 patients (105 renal tumors, 167 neuroblastomas) were included. Local models performed well on their respective center data but showed significantly reduced cross-center performance for four to seven of the nine evaluated OARs (DSC). In contrast, the FL model matched local performance for at least seven of nine OARs and achieved the best cross-center results across three metrics, with DSC gains of 0·003–0·007 over local models. FL also maintained stable performance across patient orientations and reduced false-positive kidney segmentations.

### Interpretation

Real-world FL improves cross-center robustness of CT-based OAR segmentation models in pediatric upper abdominal tumors.

### Funding

Horizon Europe/Marie Skłodowska-Curie Action co-fund project and KiTZ-Maxima Twinning Program.

## Introduction

Pediatric abdominal tumors, such as neuroblastoma and Wilms tumor, together represent roughly 12% of all pediatric malignancies.[1] Postoperative radiotherapy is indicated for a subset of these patients to improve local control.[2] Since the introduction of highly conformal target volumes often combined with intensity-modulated radiotherapy (IMRT), organs-at-risk (OARs) are usually contoured manually on computed tomography (CT),[3] which is time-consuming and introduces inter-observer variability.[4,5]

While data-driven approaches, especially deep learning (DL)–based auto-contouring, have improved efficiency and consistency in adult OARs contouring,[6–8] such initiatives remain vastly underrepresented in pediatric oncology settings[9]. This gap remains critical, as pediatric patients have marked anatomical, physiological, and imaging protocol differences compared with adults. Studies have shown that DL models, trained exclusively on adult data, can underperform in pediatric OARs.[10–12] Conversely, pediatric-specific or mixed pediatric–adult models perform better.[10–12] Furthermore, multi-observer clinical evaluations have confirmed that pediatric-specific models can significantly improve efficiency and consistency in pediatric OAR contouring.[4]

While robust DL models require representative data, pediatric oncology faces a unique data scarcity crisis; for example, pediatric cases constitute less than 1% of public medical imaging datasets.[9] Beyond the data scarcity, in-house data are fragmented across institutions. For instance, half of European pediatric oncology centers individually irradiate fewer than 25 pediatric patients per year.[13] This fragmentation necessitates multi-center collaboration for robust and reliable radiotherapy modeling, yet direct data pooling is increasingly restricted by governance constraints and privacy regulations.[14]

Federated learning (FL) enables multi-center collaboration by retaining patient data locally and sharing only model parameters, thereby maintaining data decentralization and privacy.[15] Studies have demonstrated that FL improves model robustness across various tasks, including organ segmentation,[16,17] tumor classification, and tumor segmentation.[18–20] However, most existing studies are conducted in simulated settings. Real-world implementations that provide stronger validation remain rare, accounting for only 5·2% of published FL healthcare studies.[21] The rarity and fragmentation of pediatric cancer cases make pediatric oncology an ideal field for FL. To our knowledge, only one real-world FL study has been conducted in pediatric brain tumors, focusing on classification and segmentation.[18] The application of FL to OAR contouring in pediatric upper-abdominal radiotherapy-specific workflows remains unexplored.

This study aimed to assess the feasibility and potential advantages of implementing real-world federated learning for pediatric-specific organs-at-risk segmentation across two clinical centers in the Netherlands and Germany for radiotherapy applications, serving as a proof of concept for future larger pediatric studies.

## Methods

### Dataset

This study included two international cohorts of pediatric patients with renal tumors and abdominal neuroblastoma undergoing abdominal radiotherapy: a national cohort from the Princess Máxima Center for Pediatric Oncology (PMC)/University Medical Center Utrecht (UMCU) in Utrecht, the Netherlands (UTR cohort), and a cohort from the Hopp Children's Cancer Center/Heidelberg Ion-Beam Therapy Center in Heidelberg, Germany (HEI cohort). All patients in the UTR cohort provided informed consent through the Princess Máxima Center Biobank procedure (Netherlands Trial Register NL7744; IRB approval MEC-2016-739). Patients in the HEI cohort provided informed consent under IRB approval S-377/2018.

Anonymized postoperative radiotherapy CTs and corresponding clinical delineations were retrospectively collected and processed locally. The UTR cohort, extending our previously reported cohort,[12] comprised 229 CTs from 216 pediatric patients, acquired in the supine position between 2015 and 2025. The HEI-cohort comprised 82 CTs from 57 similar patients acquired between 2017 and 2024. Within the HEI dataset, 32 CTs were acquired in a left-side lying (left-lateral) position. One CT acquired in a right-side-lying (right lateral) position was excluded.

In total, 19 (sub)structures were included in our models, comprising both clinically delineated and model-segmented structures from the two centers. The clinically delineated OARs included the left and right kidneys, left and right lungs, heart, spleen, pancreas, liver, and a combined stomach–bowel structure (three substructures), representing the organs most commonly delineated in clinical practice. Missing delineations were supplemented using a previously developed pediatric-specific model.[12] Eights OARs that were not consistently delineated in routine clinical practice were also included to expand the model's functional scope. These structures comprised the vena cava inferior (VCI), aorta abdominalis, spinal cord, vertebrae, iliopsoas muscles (L+R), and autochthonous muscles (L+R). Those segmentations were generated using the adult-based TotalSegmentator[22] and a VCI–Aorta model trained on a subset of the UTR-cohort. Further details on OAR definitions and OAR completion procedures are provided in the Supplementary Materials.

Data were partitioned per center at the patient level, with each cohort stratified to ensure representative distributions of patient orientation and IV contrast use. In the UTR-cohort, patients were split into training/validation/test sets (160/21/35), corresponding to 171/22/36 CTs. In the HEI-cohort, the corresponding splits were 35/6/15 patients and 49/8/24 CTs, respectively.

### Model and model training

All local and federated models were developed using the self-configured nnU-Net framework v2.6.0,[23] with a 3D full-resolution U-Net architecture. Region-based training was enabled for the stomach–bowel substructures. For each center, local data fingerprints were generated and aggregated on a central server to derive a unified nnU-Net plan (Figure 1), partially following a previous study.[24] All local models were then preprocessed and trained using this global plan to ensure comparability across centers. To facilitate interpretation of the model's behavior, no post-processing was applied. Different training strategies—including local versus federated models, orientation-aware augmentation (extending default rotation beyond ±30° to cover supine and left-lateral orientations), initialization strategies, and equal-weight aggregation—were explored and are reported in the Supplementary Materials. All experiments ran for 250 epochs (5 per FL round), and the final models were trained for 1000 epochs.

### Statistical Analysis

The Similarity Coefficient (DSC), Hausdorff Distance at the 95th percentile (HD95), and Mean Surface Distance (MSD) were used to assess the models' performance. For the stomach–bowel structure, all substructures were merged into a single region for evaluation. When clinical delineations were incomplete, evaluation metrics were computed only on slices with available clinical contours. Model performance was compared using the Wilcoxon signed-rank test.

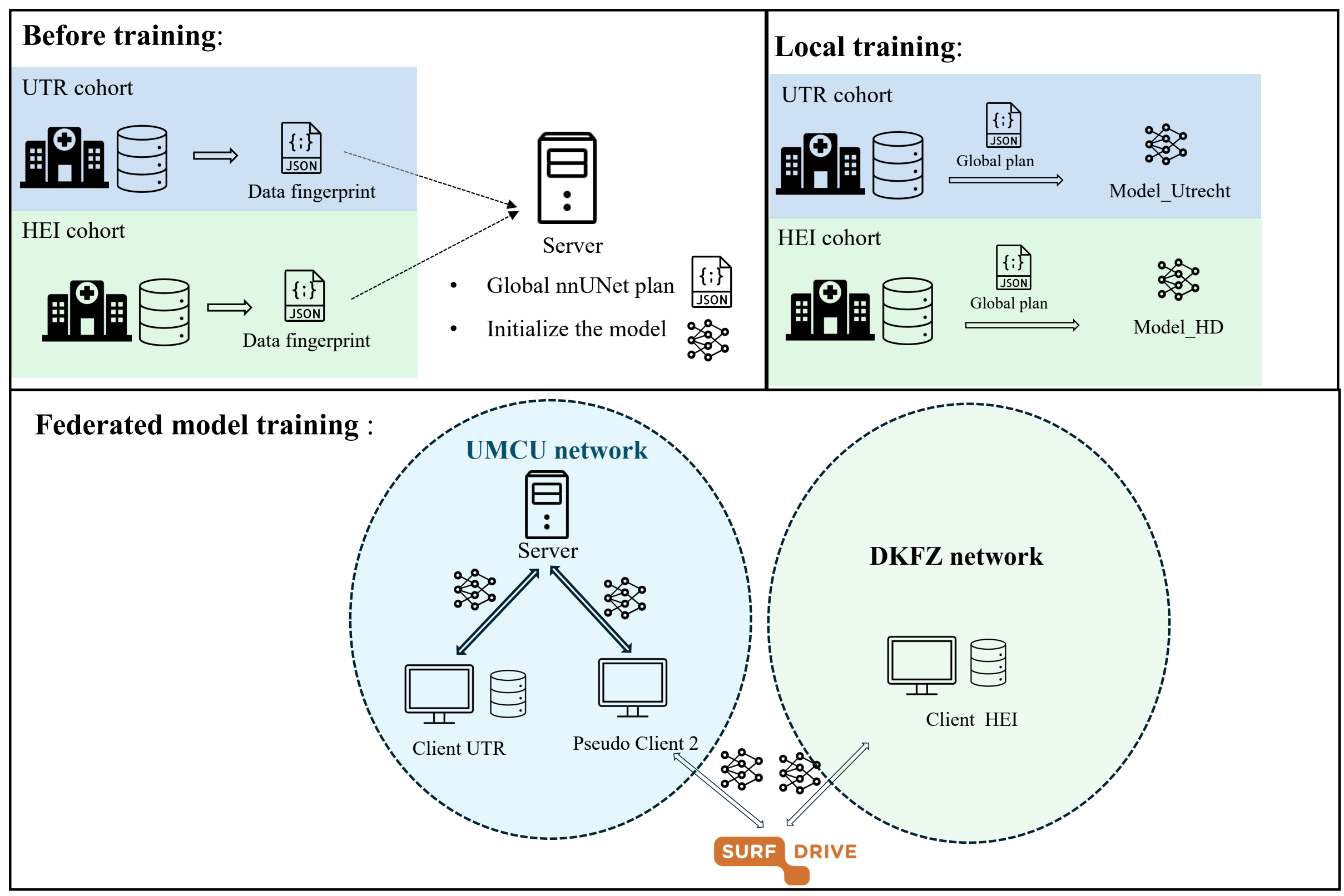

**Figure 1.** The study's workflow included data preparation, local training, and deployment of federated learning between UTR and HEI via cloud storage (SURFdrive) as an intermediate layer. Before training, the UTR and HEI cohorts generated dataset fingerprints locally from their own data and sent them to the central server to aggregate them and generate a global plan; the server also initialized the model based on this global plan. For local training, both cohorts used the same global plan to train their models, starting from the same initialized model. During federated learning, the UTR and HEI clients remained within their respective local networks, and the server was set in the UMCU network. To overcome firewall restrictions on direct communication, a pseudo-client was set up on the UMCU network, through which model weights were exchanged with the real HEI client via SURFdrive.

## Real-world federated learning deployment

The federated learning setup consisted of two clients and one central server (Figure 1), with the HEI client hosted in the DKFZ network and the UTR client and server in the UMCU infrastructure. Federated learning was implemented using NVFlare 2.5.2.[25] Due to firewall restrictions preventing direct communication between HEI and UTR, the settings model exchange was facilitated via SURFdrive, a secure cloud-based storage platform. In this setup, the HEI client uploaded model weights to SURFdrive, which a pseudo-client within the UMCU network retrieved and forwarded to the NVFlare server for aggregation. The updated global model was then returned via the pseudo-client to SURFdrive, from which the HEI client downloaded it for the subsequent training round.

The UTR-cohort ran on a system with a 96-core CPU, 754 GiB RAM, and an NVIDIA L40S GPU (46 GB), whereas the HEI-cohort used a 16-core CPU, 125 GiB RAM, and an NVIDIA Quadro RTX 6000 GPU (24 GB). The mean training time per epoch was calculated, and communication time via SURFdrive was estimated.

**Cases-based and robustness analysis**

Failure cases (DSC = 0), where the model produced no valid prediction, were reviewed and reported. To assess robustness to surgically removed organs, patients with a single remaining kidney after nephrectomy were analyzed, and false-positive kidney predictions were recorded. Model robustness was further evaluated on simulated CTs by applying manual rotations about the superior–inferior axis to represent different patient orientations: 0° (supine), 90° (right lateral), 180° (prone), and 270° (left lateral). Fully trained models were used to perform inference on these rotated CTs, and DSC was calculated for performance comparison.

**Role of the funding source**

The funder of the study was not involved in study design, data collection, data analysis, data interpretation, or writing of the report.

# Result

## Study cohort characteristics

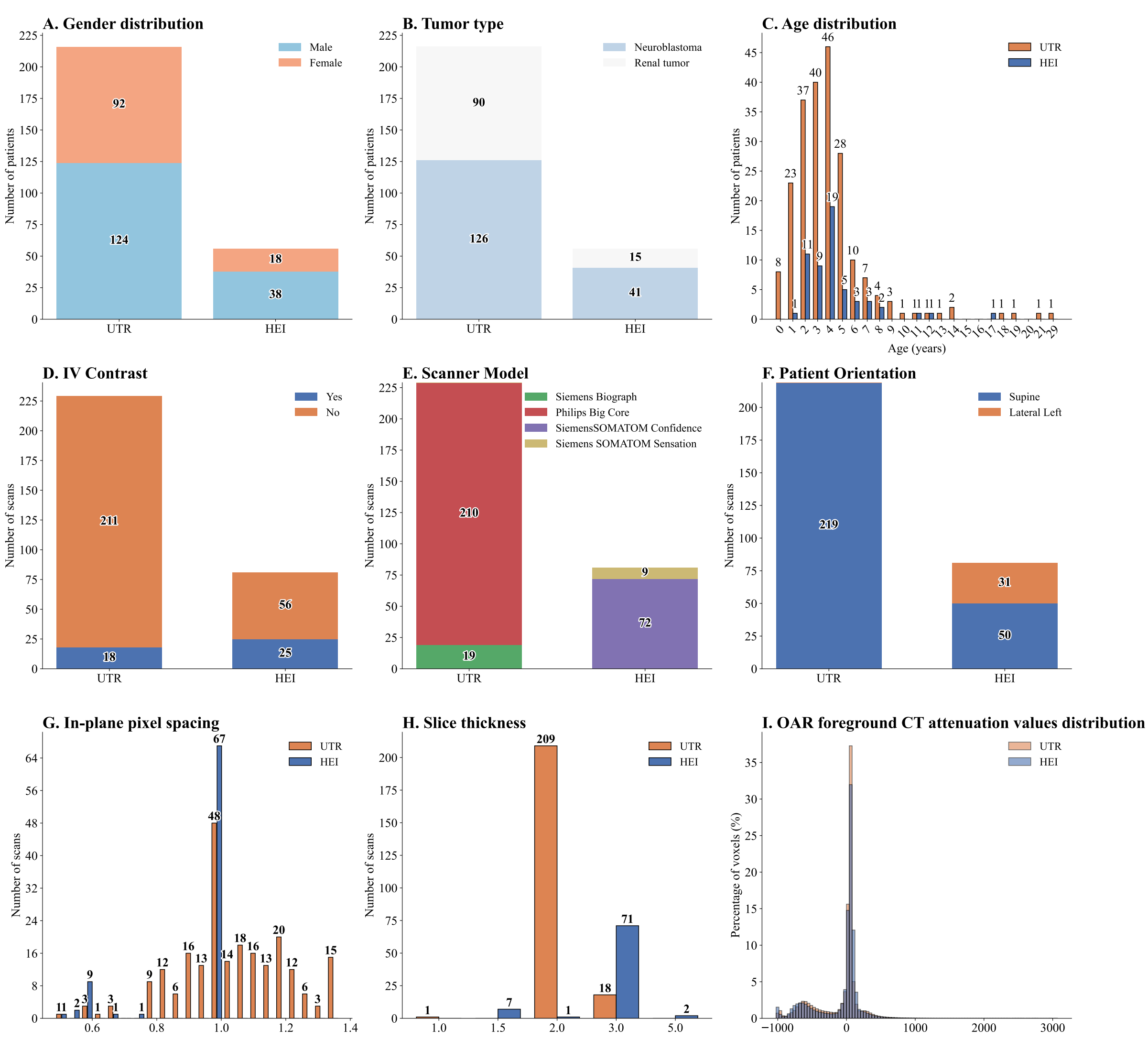

**Figure 2.** Comparison of demographic characteristics, imaging parameters, and image statistics between the UTR and HEI cohorts. Panels show (A) gender distribution, (B) tumor type, (C) age distribution, (D) IV contrast usage, (E) scanner model, (F) patient orientation, (G) in-plane pixel spacing, (H) slice thickness, and (I) OAR foreground CT attenuation values distribution (HU). All summarized information for each cohort is reported for descriptive purposes only and was not shared between centers.

In total, 310 CT scans from 272 pediatric patients across two cohorts were included in this study, comprising renal tumors (n = 105) and neuroblastoma (n = 167). The two cohorts showed a similar age distribution (Figure 2C), with most patients between 1 and 6 years of age and an overall mean age of 4.1 years (HD: 4.0±3.4 vs UTR: 4.3±2.7). The overall sex distribution was 162:110 (male:female).

Differences between the UTR and HEI cohorts were observed (Figure 2D–2F). The UTR cohort was approximately 2.8 times larger, comprising 229 CTs, whereas the HEI cohort included 81 CTs. The UTR-cohort included only supine-position CTs (vs. mixed supine and left-lateral positioning at HEI) and had a lower proportion of contrast-enhanced CTs (UTR: 18/211; HEI: 25/56). It also showed a more diverse distribution of in-plane pixel spacing and slightly smaller slice thickness (median 2 mm vs. 3 mm). Despite these differences, the foreground CT intensity distributions of the OARs largely overlapped between cohorts (Figure 2I), suggesting comparable intensity characteristics.

**Fully trained model comparison**

The final FL model configuration was selected as described in the Supplementary Materials. The UTR local model trained on 250 epochs was chosen as the initialization. Standard Federated Averaging (FedAvg) aggregation,[15] combined with orientation-aware data augmentation, was then applied for full federated training. At HEI, the mean training time per epoch across 1000 epochs was 168 s, whereas training at UTR was faster, with a mean epoch time of 82 s. Communication via SURFDrive was minimal relative to training time, estimated at approximately 4 s for both download and upload at HEI and 4 s for download and 8 s for upload at UTR. Considering five epochs per round at both centers, communication via SURFDrive accounted for less than 3% of the total training time.

Overall, the center-specific models (Model_HEI, Model_UTR) performed well on their respective training datasets but showed performance degradation when tested on the external dataset (Figure 3). In contrast, the federated model (Model_FL) demonstrated greater robustness across both cohorts.

On the HEI test dataset, Model_FL and Model_HEI showed comparable performance across most OARs, and both significantly outperformed Model_UTR on at least four OARs in terms of DSC. Between Model_FL and Model_HEI, the federated model achieved higher mean DSC 5/9 OARs (Figure S5), including a significant improvement for the right kidney (DSC +0·06). For the remaining three OARs, no significant differences were observed, with the largest decrease occurring for the left kidney, mainly driven by three challenging cases in which the organ was not predicted (DSC = 0). HD95 and MSD demonstrated small differences while maintaining trends consistent with the DSC findings (Figure S5).

On the UTR test dataset, both Model_UTR and Model_FL significantly outperformed Model_HEI. Model_FL also showed better performance than Model_HEI for 7/9 organs in terms of DSC (Figure 3). When compared directly with Model_UTR, differences were observed only for the left and right lungs, where Model_FL showed slightly lower performance; however, these differences were negligible (mean DSC difference < 0·001). In contrast, Model_FL improved performance for the stomach–bowel region, with a DSC increase of 0·19 relative to Model_UTR (Figure 4B). HD95 and MSD metrics showed similar results (Figure S6).

Across both test datasets, the federated model demonstrated improved robustness relative to local models, as reflected in mean DSC gains of 0·003–0·007 and MSD reductions of 0·158–0·390 mm and 0·697–2·418 mm (Table S6). We also observed that the UTR model, trained on a larger dataset, showed fewer OARs and significant performance degradation when evaluated on the HEI cohort, whereas the opposite was true for the smaller dataset.

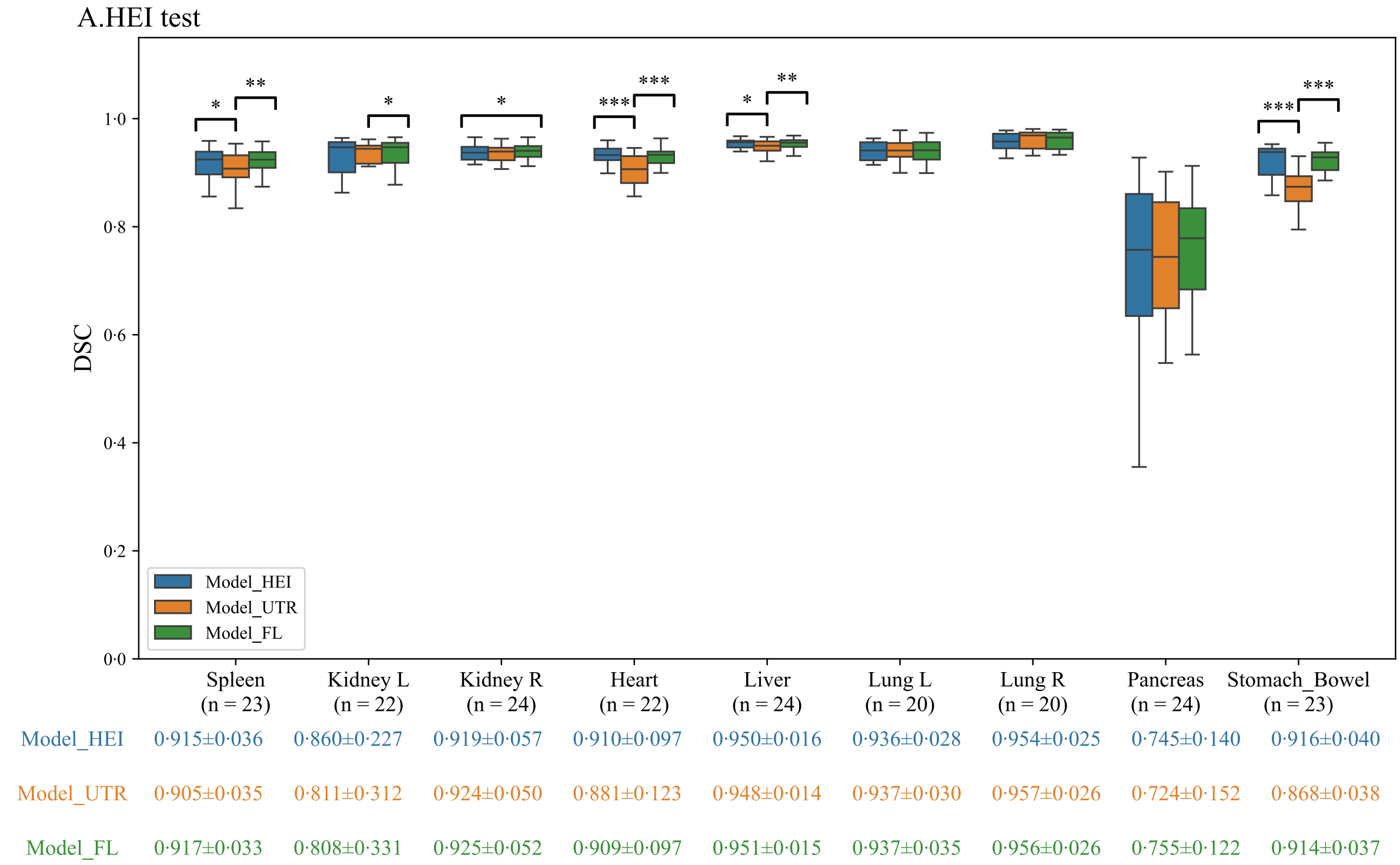

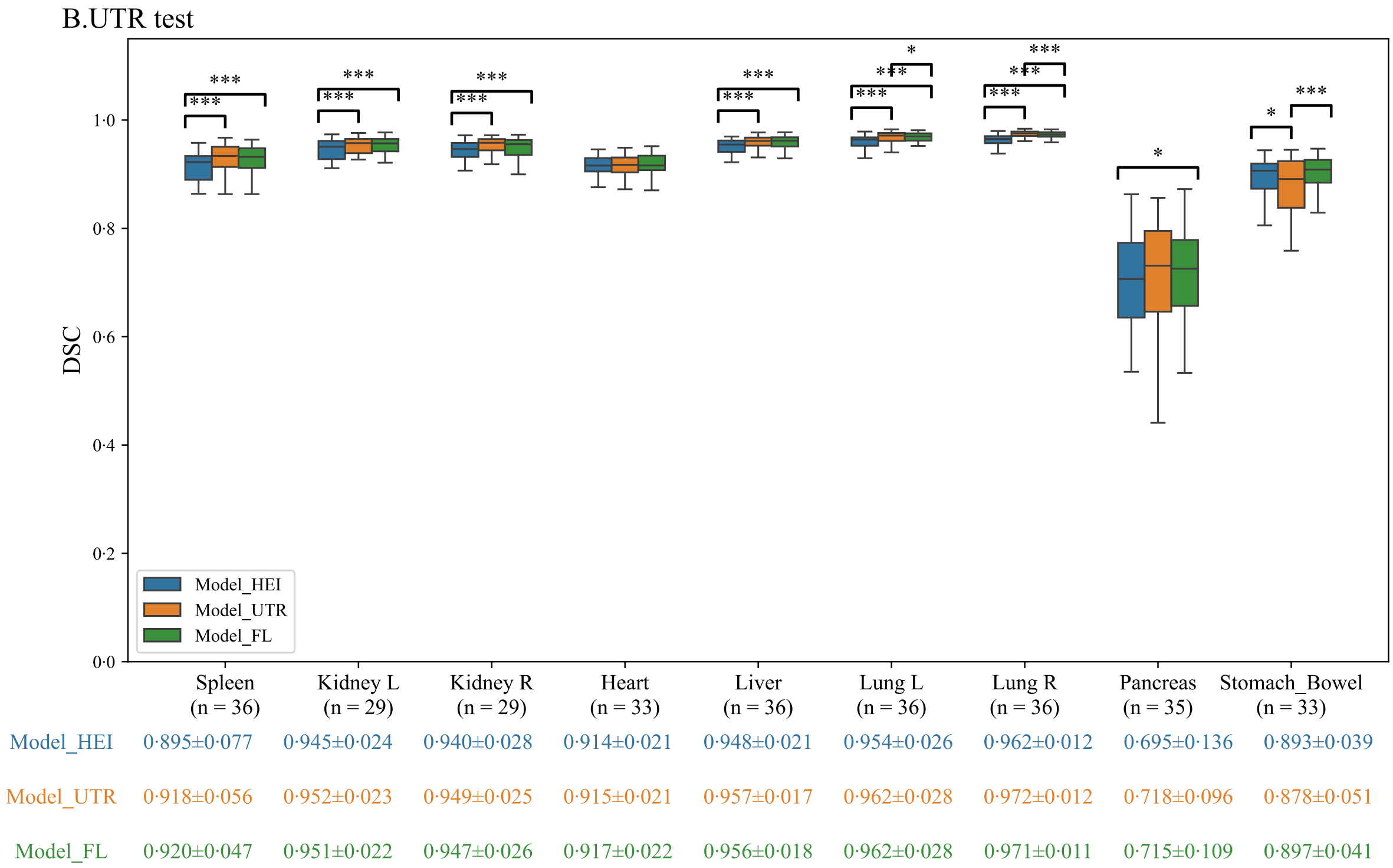

**Figure 3.** DSC of Model_FL, Model_UTR, and Model_HEI evaluated on the HEI test dataset (A) and the UTR test dataset (B). The mean and standard deviation for each model is shown below the plots. Asterisks indicate statistically significant differences based on the Wilcoxon signed rank test: **** for $p \leq 0.0001$, *** for $0.0001 < p \leq 0.001$, ** for $0.001 < p \leq 0.01$, and * for $p \leq 0.05$.

### Case-based and robust analysis

Failure cases of the federated model were noticed (Figure 4). Three cases involved abnormal presentations of the left kidney in the HEI test set (Figure 4A). In the first case, the kidney was located at an unusually low craniocaudal level, approximately at the level of the hip, and all three models misclassified this region as intestinal or bowel-bag tissue. In the second and third cases, corresponding to the same patient, CTs were acquired with and without intravenous (IV) contrast; the left kidney had minimal volume but was visible on both scans. Model_FL failed to predict the left kidney in both cases (DSC = 0) and instead misclassified it as bowel-bag tissue.

Analysis of false-positive predictions in patients with a single kidney (n = 16) showed that Model_FL produced the fewest false-positive kidney segmentations. Specifically, Model_FL generated two false positives, compared with three for Model_UTR and 10 for Model_HEI.

When evaluating model performance across different ‘simulated’ patient orientations (supine, left lateral, right lateral, and prone), Model_FL maintained the same relatively stable performance (Figure 5) as Model_HEI, as these models were exposed to all four orientations during orientation-aware data augmentation, with DSC variations within 0·016 on both datasets. In contrast, Model_UTR showed substantially reduced performance in right-lateral and prone orientations (DSC decreases > 0·3). When federated learning was performed without orientation-aware data augmentation (Table S7), using data from both centers with only supine and left-lateral orientations, the resulting model similarly failed to generalize to right-lateral and prone orientations.

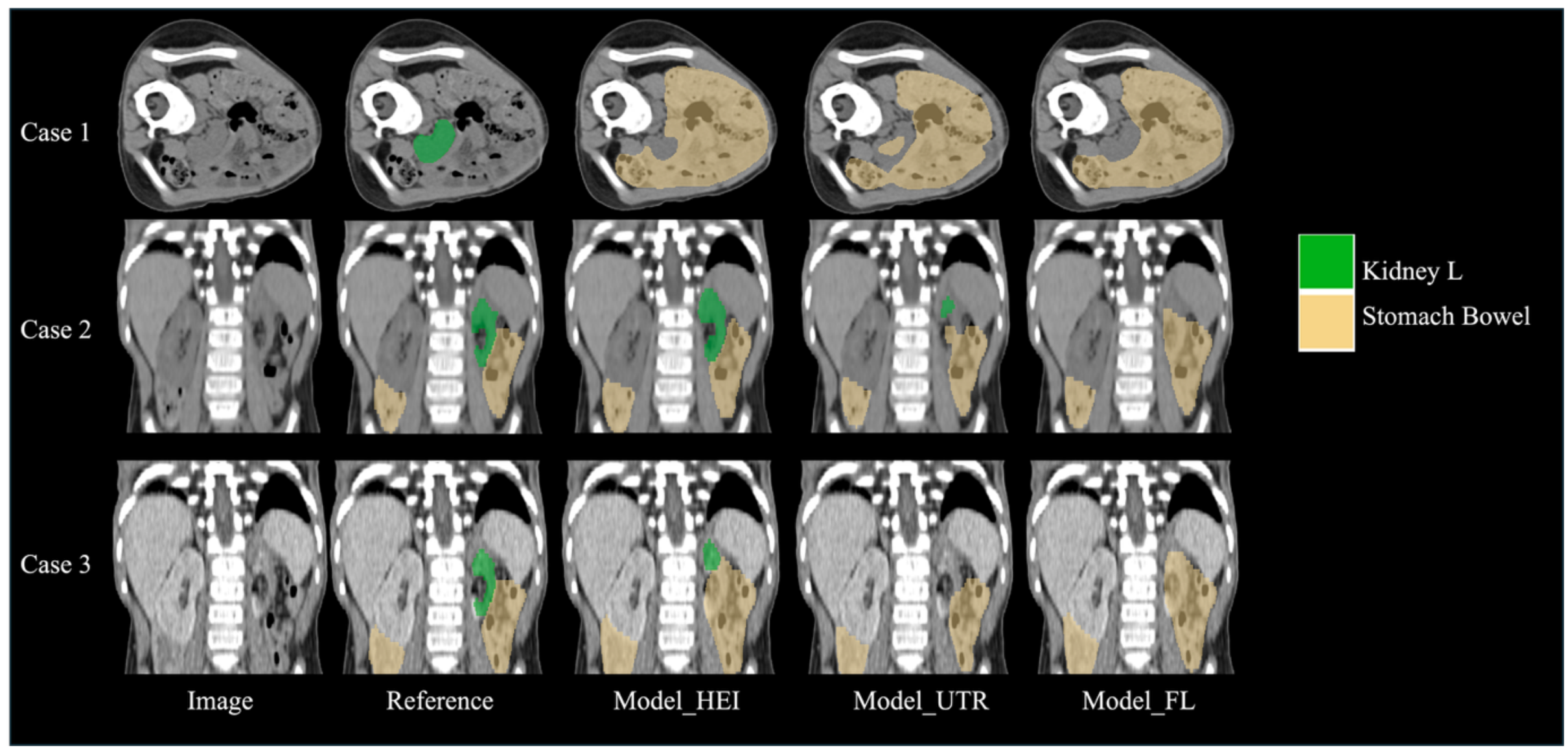

**Figure 4.** Three CTs from two patients in which the federated model failed to predict the left kidney (DSC = 0). The first row shows a CT from one patient acquired in the left lateral (left side-lying) position. The second and third rows show CTs from a second patient acquired in the supine position, without and with intravenous (IV) contrast, respectively. All images are displayed in their original patient orientations.

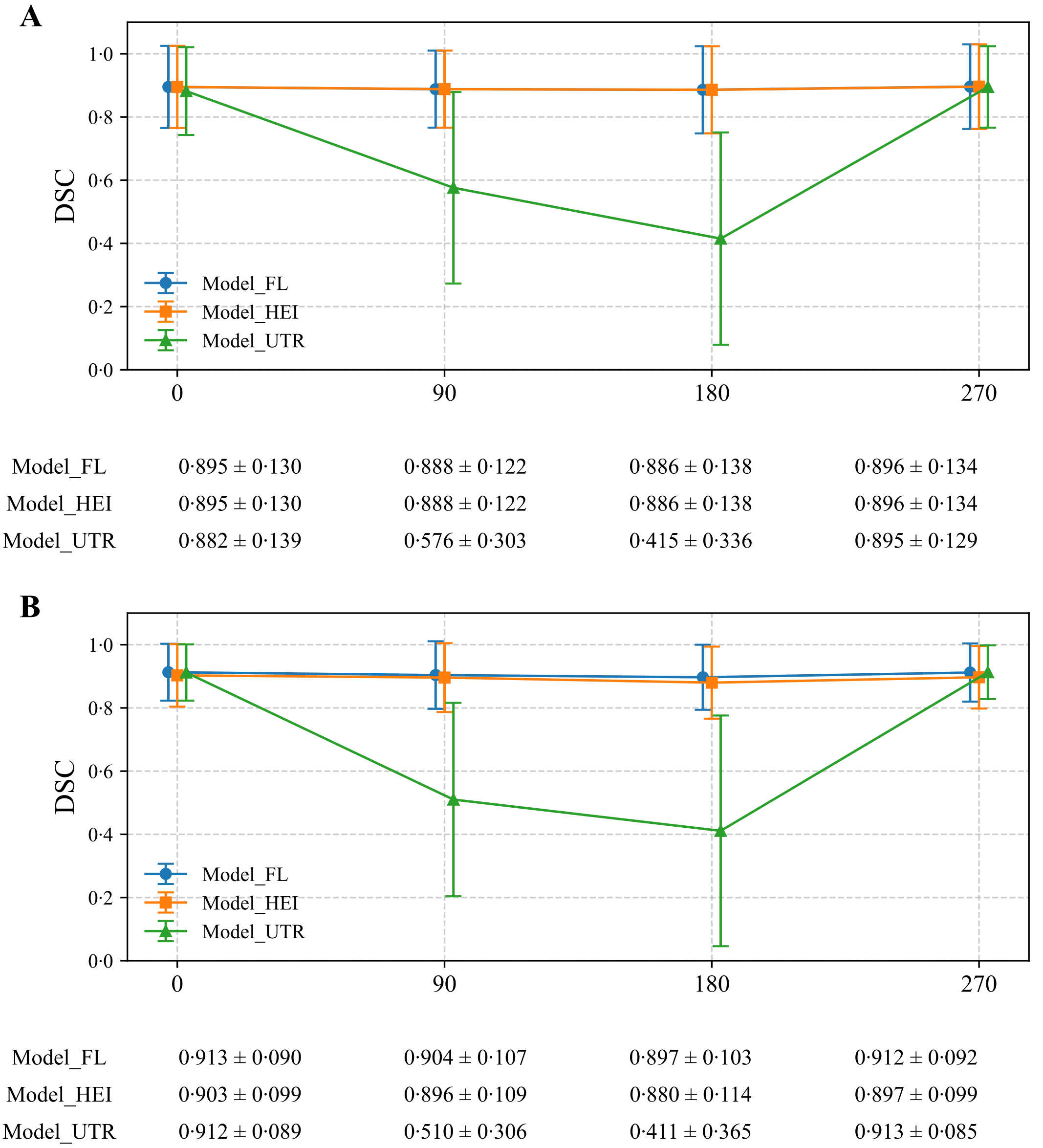

| | 0 | 90 | 180 | 270 |
|---|---|---|---|---|
| Model_FL | 0·895 ± 0·130 | 0·888 ± 0·122 | 0·886 ± 0·138 | 0·896 ± 0·134 |
| Model_HEI | 0·895 ± 0·130 | 0·888 ± 0·122 | 0·886 ± 0·138 | 0·896 ± 0·134 |
| Model_UTR | 0·882 ± 0·139 | 0·576 ± 0·303 | 0·415 ± 0·336 | 0·895 ± 0·129 |

| | 0 | 90 | 180 | 270 |
|---|---|---|---|---|
| Model_FL | 0·913 ± 0·090 | 0·904 ± 0·107 | 0·897 ± 0·103 | 0·912 ± 0·092 |
| Model_HEI | 0·903 ± 0·099 | 0·896 ± 0·109 | 0·880 ± 0·114 | 0·897 ± 0·099 |
| Model_UTR | 0·912 ± 0·089 | 0·510 ± 0·306 | 0·411 ± 0·365 | 0·913 ± 0·085 |

**Figure 5.** Performance of local models (Model_UTR and Model_HEI) and the federated model (Model_FL) on simulated CTs with different patient orientations (0° supine, 90° left lateral, 180° right lateral, and 270° prone).

## Discussion

In this study, we successfully implemented real-world FL between two European medical centers, Utrecht and Heidelberg, thereby demonstrating its feasibility. We show that FL enhances the robustness of OAR auto-contouring models for pediatric upper abdominal radiotherapy without data sharing. Our findings provide proof of concept that FL enables the development of international collaborative models to address data scarcity in rare diseases such as pediatric abdominal tumors, while improving generalization in a privacy-preserving manner.

Both local models achieved strong performance on data from their respective training centers, with results comparable to other pediatric studies across different diseases[11,26,27]. However, their performance showed a modest decline on external test datasets. In contrast, the federated model maintained stable performance across centers, with a modest robustness improvement ($\Delta$DSC = 0·003–0·007). Direct comparison with prior federated CT-based abdominal OAR studies is not feasible, as most are conducted under partial labeling scenarios.[28,29] However, an nnU-Net-based federated study reported improvements of ~0·002–0·02 for the federated model across modalities, tasks, and sites.[24] In our implementation, only model weights and a limited summary of statistics were shared between the two centers. These statistics were used to improve model design, for example, by incorporating patient orientation into orientation-aware data augmentation and using data fingerprints from both centers for global nnU-Net planning. However, federated learning can also be implemented without such prior information when data sharing is further restricted; for example, our preliminary results show that even without orientation-aware augmentation, the federated model generalizes well across centers. Using an nnU-Net plan based on the client with the largest sample size, or applying asymmetric federated averaging strategies, can avoid sharing local fingerprints.[24]

Robustness analyses showed that the federated model effectively integrated complementary strengths from the local models. The federated model was robust across all tested orientations and comparable to the HEI model, benefitting from increased orientation augmentation in the HEI-cohort. In addition, the federated model produced fewer false-positive kidney predictions in post-nephrectomy cases, comparable to the UTR model and around 5 times fewer than the HEI model (2 vs 3 vs 10). Left kidney failure outliers were observed for the federated model in the HEI test set, where kidneys with atypical positions, shapes, or volumes were not predicted.

In this study, we adapted nnU-Net with NVFlare to apply FL. The main challenge of applying was establishing secure communication between two cohorts and the server within hospital networks, where firewall restrictions prevented opening ports, as reported in previous work. [30] Involving hospital IT departments to modify firewall settings would have increased setup time and potentially raised security concerns. To avoid technical complexity and additional administrative procedures, we use a cloud storage system (SURFdrive) as an intermediate communication layer without opening a new port. Although additional communication latency increased, the upload and download times were negligible compared with model training.

This study has several limitations. First, although the FL exercise involved a targeted collaboration between two medical centers, it did not include further external validation to assess broader generalizability. Future work will involve additional centers to further evaluate the advantages of federated learning in this setting. Second, although several practical strategies were considered, we ultimately adopted the standard FedAvg approach[15] and did not explore more advanced algorithms, such as adaptive aggregation methods,[31] that may further improve performance. However, one benchmark study suggested that FedAvg remains a robust approach for medical segmentation tasks.[19] Moreover, advanced methods may yield modest performance gains while requiring extensive experimentation, thereby increasing implementation complexity and costs[17]. Third, missing OAR annotations in the HEI cohort were addressed using open-access models that were partially trained based on the UTR dataset in a previous study.[12] Although the generated segmentations had been clinically reviewed and quality-corrected, some cross-cohort influence on model generalization cannot be entirely ruled out. This reflects a common real-world challenge when working with retrospective clinical data with incomplete annotations. In the future, FL approaches incorporating knowledge distillation or semi-supervised learning[29,32] can address such inconsistencies. Finally, additional privacy-enhancing techniques, such as homomorphic encryption, may further enhance privacy.[33,34] However, these were not implemented in this study, as the collaboration was conducted within an established governance framework between the two institutions.

In conclusion, we implemented real-world federated learning for deep learning–based auto-contouring of OARs applicable to pediatric renal tumors and upper-abdominal neuroblastomas across two institutions, demonstrating that the federated models were more robust than single-center models. This study provides a proof-of-concept for privacy-preserving, multi-center collaboration in pediatric radiotherapy. It has the potential to extend federated auto-contouring models in a larger international setting, which can be attractive for rare tumors.

## Declaration of interests

We declare no competing interests.

## Acknowledgments

This work has been supported by Horizon Europe/Marie Skłodowska-Curie Action co-funded project number 101081481 (Maxima Butterfly) and by the KiTZ-Maxima Twinning Program, which stimulates collaborations and joint projects between the Hopp Children's Cancer Center (KiTZ) in Heidelberg and the Princess Máxima Center in Utrecht.

## Data sharing

The raw imaging data collected for this study will not be shared. The final federated learning model, nnU-Net configuration plans, and the implementation code for nnU-Net with NVFlare are publicly available with publication at: https://github.com/MMianyong/Fed_AbdPed.